\documentclass[aps,pra,amssymb,amsfonts,floatfix]{revtex4}
\usepackage{bm}
\usepackage{dcolumn}
\usepackage{bm}
\usepackage{mathbbol}
\usepackage{graphicx}
\usepackage{amsmath,amsthm}
\usepackage{times}
\usepackage{color}
\usepackage{lipsum}

\begin{document}

\title{Realistic many-body quantum systems vs full random matrices: static and dynamical properties}

\author{Eduardo Jonathan Torres-Herrera $^{1}$, Jonathan Karp $^{2}$, Marco T\'avora $^{2}$ and Lea F. Santos $^{2}$}

\affiliation{$^1$Instituto de F{\'i}sica, Universidad Aut\'onoma de Puebla, Apt. Postal J-48, Puebla, Puebla, 72570, Mexico}
\affiliation{$^2$Department of Physics, Yeshiva University, New York, New York 10016, USA}

\begin{abstract}
We study the static and dynamical properties of isolated many-body quantum systems and compare them with the results for full random matrices. In doing so, we link concepts from quantum information theory with those from quantum chaos. In particular, we relate the von Neumann entanglement entropy with the Shannon information entropy and discuss their relevance for the analysis of the degree of complexity of the eigenstates, the behavior of the system at different time scales and the conditions for thermalization. A main advantage of full random matrices is that they enable the derivation of analytical expressions that agree extremely well with the numerics and provide bounds for realistic many-body quantum systems.
\end{abstract}

\maketitle

\section{Introduction}

Advances in quantum information science and many-body quantum physics have been closely intertwined. Quantum computers, for example, are many-body quantum systems. To build the first, one needs a better understanding of the latter. At the same time, one of the main reasons for developing quantum computers is to simulate many-body quantum systems~\cite{PreskillProceed}. New numerical methods, such as the density matrix renormalization group employed in the studies of many-body quantum systems \cite{White1992,Feiguin2005a}, rely on notions of entanglement. In particular, the accuracy of these methods requires a limited growth of entanglement with time and system size~\cite{Eisert2010}. An important task in quantum information processing is the faithful transfer of quantum states. To accomplish this, one needs a precise characterization of the dynamics of many-body quantum systems at different time scales. In turn, studies of nonequilibrium quantum dynamics often touch upon the old quest of deriving statistical mechanics and thermodynamics from first principles.

Thermalization in isolated many-body quantum systems is caused by strong interactions between their particles (or quasiparticles) and is intimately related to the notion of quantum chaos~\cite{ZelevinskyRep1996,Borgonovi2016,AlessioARXIV}. Quantum chaos refers to signatures that one finds at the quantum level, such as level repulsion, that tell us whether or not the classical counterpart of the system is chaotic~\cite{Guhr1998}. A main feature of a classically-chaotic system is the extreme sensitivity of its dynamics to the initial conditions. At the quantum level, one can no longer talk about phase-space trajectories, but one can still expect to find quantum signatures of classical chaos, such as those associated with spectrum properties~\cite{Gubin2012}. This notion of quantum chaos has, however, been extended to systems that may not even have a classical limit. In addition to the properties of the spectrum, quantum chaos is also directly connected with the emergence of chaotic eigenstates, that is highly delocalized states that are similar to (pseudo)-random vectors~\cite{ZelevinskyRep1996,Borgonovi2016}. The level of delocalization of the eigenstates is measured with quantities, such as the participation ratio and the Shannon information entropy~\cite{ZelevinskyRep1996,Borgonovi2016,Gubin2012}.

In this work, we explore the relationship between measures of entanglement and measures of delocalization~\cite{SantosEscobar2004,Monasterio2005b,Lakshminarayan2005,Brown2008,Dukesz2009,Giraud2009}. We show that the von Neumann entanglement entropy and the Shannon information entropy provide similar information about the system. The first requires the partial trace of the density matrix, being computationally more expensive than the second, which deals with the states of the entire system. 

We also analyze the evolution of many-body quantum systems from very short times until the moment they reach equilibrium. Our analysis is based on the behavior of the survival probability (probability of finding the initial state later in time), the Shannon entropy and the entanglement entropy. We show that the short- and intermediate-time evolutions are generic when the system is taken far from equilibrium, while the long-time dynamics depends on the level of chaoticity of the initial state. At intermediate times, the decay of the survival probability may be faster than exponential~\cite{Izrailev2006,Torres2014PRA,Torres2014NJP,TorresProceed,Torres2014PRAb,Torres2014PRE,TorresKollmar2016,Torres2015,Torres2016BJP}. At long times, the survival probability necessarily shows a power law decay~\cite{TavoraARXIV}. From the value of the power law exponent, one may anticipate whether or not the initial state 
will~thermalize.

We start our studies in Section~\ref{section2} using full random matrices. These are matrices filled with random numbers. Early connections between the properties of the spectrum of quantum systems and classical chaos were made in the context of full random matrices and became known as the Bohigas--Giannoni--Schmit conjecture~\cite{Bohigas1984}. Full random matrices are not the most suitable models for many-body quantum systems, because they imply that all of the particles interact at the same~time. Yet, they allow for the derivations of analytical results that can serve as references and bounds for the analysis of realistic models. In Section~\ref{section3}, we then investigate the static properties and the dynamics of realistic many-body quantum systems described by one-dimensional spin-1/2 models. We consider both integrable and chaotic limits. These models are similar to those employed in experiments that use cold atoms to study nonequilibrium dynamics~\cite{Trotzky2008,Trotzky2012} 
and~thermalization~\cite{kinoshita06,Kaufman2016}.


\section{Full Random Matrices and Thermalization}\label{section2}

Full random matrices are matrices filled with random numbers whose only constraint is to satisfy the symmetries of the system they are trying to describe. They were used extensively by Wigner to model the spectra of heavy nuclei, which are very complex systems~\cite{Wigner1958}. In this approach, interactions are treated statistically, and their details are overlooked. This simple idea led to results that agreed very well with data from real nuclei and was soon employed in the analysis of other complex systems, such as atoms, molecules and quantum dots~\cite{Guhr1998}. We discuss some of the static properties of these matrices in Sections~\ref{section2.1} and \ref{section2.2}, and some of their dynamic properties in Section~\ref{section2.3}. In Section~\ref{section2.4}, we discuss why thermalization is trivially achieved for full random matrices.

\subsection{Eigenvalues: Density of States and Level Repulsion}\label{section2.1}

There are different kinds of ensembles of full random matrices defined according to the symmetries that the matrices satisfy~\cite{Dyson1962,MehtaBook,Guhr1998}. When modeling systems with time reversal symmetry, random matrices of the Gaussian orthogonal ensemble (GOE) are used. These are ${\cal D} \times {\cal D} $ real and symmetric matrices with entries from a Gaussian distribution with mean zero,
\begin{eqnarray}
&&\langle H_{ij}^2 \rangle =\langle H_{ji}^2 \rangle = 1 \hspace{0.3 cm} \text{for} \hspace{0.3 cm} i<j \nonumber \\
&&\langle H_{ii}^2 \rangle = 2 .
\end{eqnarray}

In practice, GOE full random matrices can be obtained by generating a matrix with ${\cal D}^2$ random numbers and then adding it to its transpose. The density of states of full random matrices follows the standard semicircle distribution~\cite{Wigner1955},
\begin{equation}
\rho^{DOS} (E)= \frac{2}{\pi {\cal E}} \sqrt{1- \left( \frac{E}{{\cal E}} \right)^2 },
\label{DOSfrm}
\end{equation}
where $2 {\cal E}$ is the length of the spectrum, that is $-{\cal E} \leq E \leq {\cal E}$. 

A key property of full random matrices and a main feature of quantum chaos is the strong repulsion between neighboring levels, as captured, for example, by the nearest-neighbor level spacing distribution $P(s)$, where $s$ is the spacing between neighboring rescaled energies. The unfolding procedure~\cite{Gubin2012} guarantees that the mean level spacing of the rescaled eigenvalues is one. For the unfolded spectrum of GOE matrices, one finds:
\begin{equation}
P(s)= \frac{\pi s}{2} \exp \left(- \frac{\pi s^2}{4} \right).
\end{equation}

\begin{figure}[ht]
\centering
\includegraphics*[width=8.3cm]{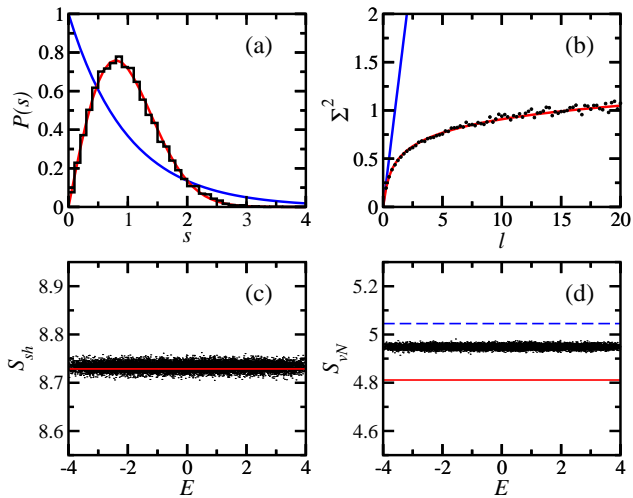}
\caption{We consider a Gaussian orthogonal ensemble (GOE) full random matrix. Top panels: level spacing distribution (\textbf{a}) and level number variance (\textbf{b}). Bottom panels: Shannon information entropy (\textbf{c}) and von Neumann entanglement entropy (\textbf{d}) for all eigenstates. Horizontal solid lines give $\ln(0.48 {\cal D})$ in (c) and $\ln(0.48 {\cal D}_{A})$ in (d). The horizontal dashed line in (d) corresponds to $S_{vN}^{rand}=\ln {\cal D}_{A} - 1/2$. The random numbers of the full random matrix are rescaled so that ${\cal E} \sim 4$. We choose ${\cal D}=16!/8!^2=12870$ and ${\cal D_A}=2^8=256$ in analogy with the matrix sizes used in Section~\ref{section3}.}
\label{Fig:FRMstatic} 
\end{figure}

This contrasts with the level spacing distribution of a sequence of uncorrelated eigenvalues, where the levels are not prohibited from crossing and the distribution is Poisson, $P(s)=\exp(-s)$. Level repulsion causes the rigidity of the spectrum. As a result, the variance $\Sigma^2 (\ell) $ of the number of unfolded eigenvalues in an interval of length $\ell$ grows logarithmically with $\ell$. For the GOE, one has:
\begin{equation}
\Sigma^2 (\ell) = \frac{2}{\pi^2} \left( \ln(2 \pi \ell) + \gamma_e + 1 -\frac{\pi^2}{8} \right) ,
\end{equation}
where {$\gamma_e = 0.5772\ldots $ }is Euler's constant. The level number variance of full random matrices is between the linear dependence $ \Sigma^2 (\ell) =\ell$ found for uncorrelated eigenvalues and $ \Sigma^2 (\ell) =0$ reached by the completely rigid spectrum of the harmonic oscillator. $P(s)$ and $\Sigma^2 (\ell) $ are complementary. The former characterizes the short-range fluctuations of the spectrum, and the latter characterizes the long-range fluctuations. Both are shown in Figures~\ref{Fig:FRMstatic}a and \ref{Fig:FRMstatic}b, respectively.

\subsection{Eigenstates: Delocalization and Entanglement Measures}\label{section2.2}

There are various ways to quantify how much a state spreads out in a certain basis. One of them is the participation ratio $PR$. Given an eigenstate $|\psi_{\alpha} \rangle = \sum_k C_k^{\alpha} |\phi_k\rangle$ written in a basis $|\phi_k\rangle$, 
$PR^{(\alpha)} = 1/\sum_k |C_k^{\alpha}|^4$. A comparable quantity is the Shannon information entropy, defined as:
\begin{equation}
S_{Sh}^{(\alpha)} = - \sum_k |C_k^{\alpha}|^2 \ln |C_k^{\alpha}|^2.
\end{equation}
The values of $PR^{(\alpha)}$ and $S_{Sh}^{(\alpha)}$ depend on the chosen basis. In the case of full random matrices, where one deals with ensembles of matrices filled with random numbers, the notion of basis is not well defined, but we can still associate values with the delocalization measures. 

All eigenstates of full random matrices are (pseudo)-random vectors. In the particular case of GOEs, the coefficients are real random numbers from a Gaussian distribution satisfying the normalization condition. All eigenstates are therefore equivalent and lead to approximately the same values of the participation ratio, $PR^{GOE}\sim {\cal D}/3$, and of the Shannon entropy: 
\begin{equation}
S_{Sh}^{GOE} \sim \ln (0.48 {\cal D}) .
\label{ShGOE}
\end{equation}
The latter is shown in Figure~\ref{Fig:FRMstatic}c. This result can be derived as follows. The sum is approximated by an integral, $\sum_k f(C_k) \rightarrow {\cal D} \int_{-\infty}^{\infty} f(C) P(C) dC$, where $P(C) = \sqrt {{\cal D}/2\pi } {{\mathop{\rm e}\nolimits} ^{ - {\cal D}{C^2}/2}}$ is the Gaussian probability distribution of the components. This approach guarantees that the average of the coefficients is $\overline{C} =0$, the average of their squares is $\overline{C^2} =1/{\cal D}$, and $\sum_k |C_k^{\alpha}|^2 \rightarrow {\cal D} \int_{-\infty}^{\infty} C^2 P(C) dC=1$. The entropy is therefore:
\begin{equation}
S_{Sh}^{GOE} \sim - {\cal D} \sqrt{\frac{{\cal D}}{2 \pi}} \int_{-\infty}^{\infty} \exp \left( - \frac{{\cal D} C^2}{2} \right) C^2 \ln C^2 dC= 
-2 + \ln 2 + \gamma_e + \ln{\cal D} \sim \ln (0.48 {\cal D}) .
\end{equation}
Notice that the Shannon entropy of random vectors is smaller than the maximum value $S_{Sh}^{max}=\ln {\cal D}$ obtained when all weights $|C_k^{\alpha}|^2=1/{\cal D}$.

Similarly, there are various methods for quantifying the amount of entanglement in a state. We focus on the von Neumann entanglement entropy, $S_{vN}$ \cite{Amico2008}. Its computation requires the bipartition of the system in subsystems A and B and the partial trace of one of the two. $S_{vN}(\rho_A)$ is the von Neumann entropy of the reduced density matrix $\rho_A = \text{Tr}_B \rho$, where $\rho$ is the density matrix associated with the total system. It gives:
\begin{equation}
S_{vN}= - \text{Tr} \left( \rho_A \ln \rho_A \right) .
\end{equation}

Maximum entanglement occurs when the reduced density matrix is maximally mixed, that is when it is the normalized identity matrix. In this case, $S_{vN}^{max}\sim \ln {\cal D}_{A}$, where ${\cal D}_{A}$ is the dimension of~$\rho_A$. 
Studies have shown that this limit is nearly achieved by random pure states~\cite{Lubkin1978,Page1993}. In particular, Page~\cite{Page1993} obtained $S_{vN}^{rand} \simeq \ln {\cal D}_{A} - {\cal D}_{A}/(2 {\cal D}_{B})$ for a pure random state of dimension ${\cal D}={\cal D}_{A} {\cal D}_{B}$ and ${\cal D}_{A} \leq {\cal D}_{B}$. Here, we consider instead that ${\cal D}_{A} {\cal D}_{B}>{\cal D}$, with ${\cal D}_{A} = {\cal D}_{B}$. Because of this choice, $S_{vN}^{rand}$ (dashed line in Figure~\ref{Fig:FRMstatic}d) is slightly above our numerical results for the entanglement entropy. The values for ${\cal D}$ and ${\cal D}_A$ that we select are motivated by the comparison that we later make in Section~\ref{section3} with spin systems that have matrices of these same dimensions. We also find numerically that the results for the entanglement entropy are slightly larger than $\ln(0.48 {\cal D}_{A})$ (solid line Figure~\ref{Fig:FRMstatic}d). However, since these differences are minor, when ${\cal D}_A$ is large, we chose to write, in analogy with $S_{Sh}^{GOE}$,~that: 
\begin{equation}
S_{vN}^{GOE} \sim \ln(0.48 {\cal D}_{A}) .
\label{SvGOE}
\end{equation}

\subsection{Time Evolution: Entropy Growth and Survival Probability}\label{section2.3}

To study the evolution under a GOE full random matrix, we assume that a fictitious basis of product states $|\phi_{k} \rangle$ was used to construct the matrix and that the initial state $|\Psi(0) \rangle = |\phi_{ini} \rangle$ is the basis vector in the middle of the matrix, $k=ini={\cal D}/2$. The entanglement entropy at $t=0$ is therefore $S_{vN}(0) =0$, but it grows as the state evolves. 

Written in the energy eigenbasis, the initial state is:
\begin{equation}
|\Psi(0) \rangle = \sum_{\alpha=1}^{{\cal D}} C^{\alpha}_{{\cal D}/2} |\psi_{\alpha}\rangle.
\label{PSIo}
\end{equation}

Since the eigenstates $|\psi_{\alpha}\rangle$ are random vectors, so is $|\Psi(0) \rangle$. The Shannon entropy of $|\Psi(0) \rangle$ in the energy eigenbasis (also known as diagonal entropy; see Equation~(\ref{Sdiagonal})) gives values very similar to what we find in Figure~\ref{Fig:FRMstatic}, $S_{Sh}^{ini} = - \sum_{\alpha=1}^{{\cal D}} | C^{\alpha}_{{\cal D}/2} |^2 \ln | C^{\alpha}_{{\cal D}/2} |^2 \sim \ln(0.48 {\cal D})$. We note, however, that to study dynamics, we consider the spreading in time of $|\phi_{ini} \rangle$ over the other basis vectors. In Figure~\ref{Fig:FRM}a, we show the evolution of the Shannon entropy given by:
\begin{equation}
S_{Sh}(t)= - \sum_{k=1}^{ {\cal D} } W_k (t) \ln W_k (t) , 
\label{SforFRM}
\end{equation}
where the probability $W_k(t)$ for the initial state to be found in state $|\phi_k \rangle$ at time $t$ is:
\begin{eqnarray}
&& W_k (t) = \left| \langle \phi _k | e^{ - iHt} | \phi _{ini} \rangle \right|^2 = \left| \sum_{\alpha} C_k^{\alpha *} C_{ini}^{\alpha} e^{ - i E_{\alpha} t} \right|^2 = \left| \int P_{k,ini} (E) e^{ - iEt} dE \right|^2, \hspace{0.2 cm} \text{where} \nonumber\\
&& P_{k,ini} (E) = \sum_{\alpha} C_k^{\alpha *} C_{ini}^{\alpha} \delta (E - E_{\alpha }).
\label{Pkini}
\end{eqnarray}
and Planck's constant is one. We can obtain $W_k(t)$ by taking the Fourier transform of the distribution $P_{k,ini}(E) $, if the latter is known. Alternatively, a simpler approach is to separate the contributions to $W_k(t)$ into those coming from $k=ini$ and those from $k\neq ini$, as we do next.

\subsubsection{Survival Probability and Power Law Decays}\label{section2.3.1}

$W_{ini}(t)$ is the survival probability of the initial state. It corresponds to the probability of finding the initial state $ | \Psi(0) \rangle = |\phi_{k=ini} \rangle$ later in time,
\begin{equation}
W_{ini}(t) \equiv \left| \langle \Psi(0) | e^{-i H t} | \Psi(0) \rangle \right|^2 
= \left|\sum_{\alpha} |C^{\alpha}_{ini} |^2 e^{-i E_{\alpha} t} \right|^2 = \left| \int P_{ini,ini}(E) e^{-i E t} dE \right|^2. 
\end{equation}

Above, $P_{ini,ini}(E) =\sum_{\alpha} \left| C_{ini}^{\alpha} \right|^2 \delta (E - E_{\alpha }) $ is the energy distribution of the initial state weighted by the components $|C^{\alpha}_{ini} |^2$. It is often referred to as the local density of states (LDOS) or strength function. For full random matrices, this distribution is a semicircle~\cite{Torres2014PRA,Torres2014NJP}, as the density of states in Equation~(\ref{DOSfrm}). This is seen in Figure~\ref{Fig:FRM}a. The Fourier transform of a semicircle gives:
\begin{equation}
W_{ini}^{GOE}(t) = \frac{[{\cal J}_1( 2 \sigma_{ini} t)]^2}{\sigma_{ini}^2 t^2},
\label{FforFRM}
\end{equation}
 where ${\cal J}_1$ is the Bessel function of the first kind,
 \begin{eqnarray}
\sigma_{ini}^2 &=& \langle \Psi(0) |H^2 |\Psi(0) \rangle - \langle \Psi(0) |H |\Psi(0) \rangle^2 \nonumber \\
&=&
 \sum_{\alpha} |C^{\alpha}_{ini} |^2 E_{\alpha}^2 - \left( \sum_{\alpha} |C^{\alpha}_{ini} |^2 E_{\alpha} \right)^2 =\sum_{\alpha} |C^{\alpha}_{ini} |^2 (E_{\alpha} - E_{ini})^2 \nonumber \\
&=& \sum_k \langle \Psi(0) |H| \phi_k \rangle \langle \phi_k|H|\Psi(0) \rangle - \langle \Psi(0) |H |\Psi(0) \rangle^2 =
\sum_{k \neq ini} \left| \langle \phi_k|H|\Psi(0) \rangle \right|^2
 \label{sigma} 
 \end{eqnarray}
 is the variance of the energy distribution of the initial state and: 
 \begin{equation}
 E_{ini} = \langle \Psi(0) |H |\Psi(0) \rangle =\sum_{\alpha} |C^{\alpha}_{ini} |^2 E_{\alpha} \
\end{equation}
 is the energy of $| \Psi(0) \rangle$. For the initial state considered here, $\sigma_{ini}^{GOE} \sim {\cal E}/2$ and $E_{ini}^{GOE} \sim 0$. The~agreement between the numerical results for $W_{ini}(t)$ and Equation~(\ref{FforFRM}) is excellent, as seen in Figure~\ref{Fig:FRM}b.
 
The asymptotic behavior of $W_{ini}^{GOE}(t)$ for $t\gg \sigma_{ini}^{-1}$ is a power law and $\propto t^{-3}$ \cite{TorresKollmar2016,TavoraARXIV}. This can be derived from a theorem proven in~\cite{Erdelyi1956}. For the semicircle, $P^{GOE}_{ini,ini}(E) = ( {\cal E} + E)^{\xi} \eta (E)$ with $\xi=1/2$ and $\eta(E) = 2({\cal E} - E)^{1/2}/(\pi {\cal E}^2)$. The theorem says that if $0<\xi <1$ and $\eta(E)$ is $N$ times continuously differentiable for $- {\cal E} \leq E < {\cal E}$, then for $t\rightarrow \infty$, we have~\cite{Erdelyi1956,Urbanowski2009}:
\begin{equation}
\int_{- {\cal E}} ^{\cal E} P_{ini,ini}(E) e^{-i E t} dE = - i \xi e^{i{\cal E}t} \sum_{n = 0}^{N - 1} \frac{\Gamma (n + \xi )}{n!} e^{ - i\pi (n + \xi + 2)/2} \eta_0^{(n)} t^{ - n - \xi - 1} + O( t^{ - N}),
\end{equation}
where $\eta_0^{(j)} = \lim_{E \rightarrow -\epsilon^{+}} \eta^{(j)} (E)$ and $j=0, 1, \ldots N$.
The dominant term for $n=0$ leads to: 
\begin{equation}
W_{ini}^{GOE}(t \gg \sigma_{ini}^{-1}) \propto t^{-3}.
\label{Eq:powerGOE}
\end{equation}

This inverse power law decay is a manifestation of the Khalfin effect~\cite{Khalfin1958}, which refers to the slow decay of the survival probability at long times due to the unavoidable bounds (at least the ever present lower bound) in the spectrum of any quantum system. Khalfin showed that the usual exponential decay of unstable states could not persist for long times. Later studies showed that the decay should become a power law~\cite{MugaBook,Campo2011,Urbanowski2009,TavoraARXIV} with the exponent depending on the properties of $P_{ini,ini}(E) $ \cite{TavoraARXIV,Torres2016BJP,Torres2015}.

\subsubsection{Entropy Growth}

\begin{figure}[ht]
\centering
\includegraphics*[width=11cm]{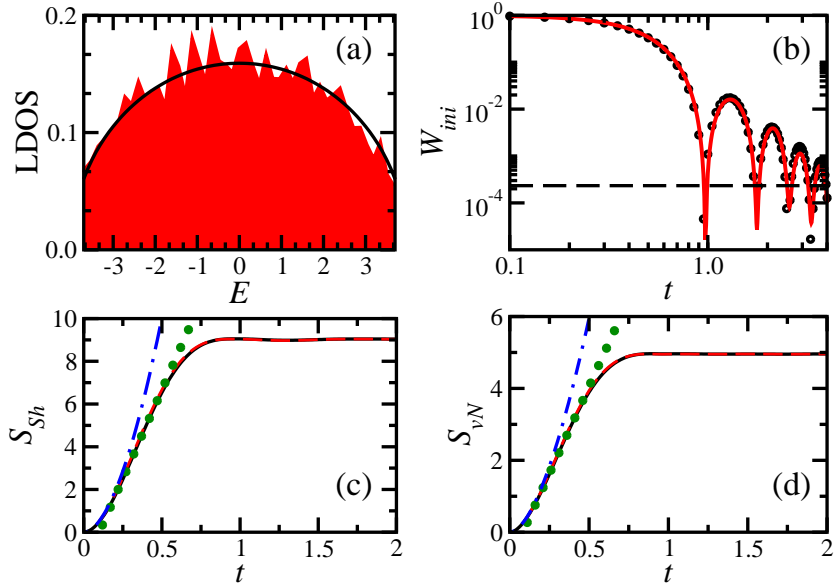}
\caption{ We consider the initial state (\ref{PSIo}) evolving under a GOE full random matrix. Top panels: local density of states (LDOS) (\textbf{a}) and the survival probability (\textbf{b}). The symbols in (b) refer to Equation~(\ref{FforFRM}), and the dashed horizontal line is $\overline{W}_{ini}^{GOE}\sim 3/{\cal D}$ from Equation~(\ref{Eq:satFgoe}).
Bottom panels: evolution of the Shannon entropy (\textbf{c}) and von Neumann entanglement entropy (\textbf{d}). Dashed lines correspond to the expression in Equation~(\ref{ShNpc}) with $N_{pc}= \langle \exp[S_{Sh} (t)]\rangle$ in (c) and $N_{pc}= \langle \exp[S_{vN} (t)]\rangle$ in (d). The dashed lines are indistinguishable from the numerical results (solid lines). Dot-dashed lines are Equation~(\ref{Eq:shortGOE}). Circles represent linear fits, $S_{Sh, vN} =a_{Sh, vN} +b_{Sh,vN} t$, with $a_{Sh}=-1.65, a_{vN}=-0.87$ and $b_{Sh}=16.6, b_{vN}=9.97$.~The random numbers of the full random matrix are rescaled so that ${\cal E} \sim 4$; ${\sigma _{ini}} = 2$. We choose ${\cal D}=12870$ and ${\cal D_A}=256$ as in Figure~\ref{Fig:FRMstatic}.}
\label{Fig:FRM} 
\end{figure}

We now come to Equation~(\ref{SforFRM}) and rewrite it as~\cite{Flambaum2001b,Santos2012PRL,Santos2012PRE}:
\begin{eqnarray}
S_{Sh}(t) &=& - W_{ini}(t) \ln W_{ini}(t) - \sum_{k\neq ini}^{ {\cal D} } W_k (t) \ln W_k (t) \nonumber \\
&\sim & - W_{ini}(t) \ln W_{ini}(t) - [1- W_{ini}(t)] \ln \left[ \frac{1- W_{ini}(t)}{N_{pc}} \right], 
\label{ShNpc}
\end{eqnarray}
where the use of $1- W_{ini}(t)$ in the second line is motivated by the normalization condition $\sum_{k=1}^{ {\cal D} } W_k (t) =1$ and the ratio $[1- W_{ini}(t)]/N_{pc}$ by the fact that at very long times, $W_{ini}(t)$ should be very small and $S_{Sh}\rightarrow \ln N_{pc}$. $N_{pc}$ refers to the number of states that contribute to the evolution of the initial state. In the case of full random matrices, we would expect $N_{pc}^{GOE} \sim 0.48 {\cal D}$. However, by doing an average of the numerical values of $\exp[S_{Sh} (t)]$ at long times, after the saturation of the evolution, we actually find that $N_{pc}$ is slightly larger than $0.48 {\cal D}$. Why this is so still needs to be understood. The semi-analytical expression above, with the result for $N_{pc} \sim \langle \exp[S_{Sh} (t)] \rangle $, agrees extremely well with the numerical results for the Shannon entropy (see Figure~\ref{Fig:FRM}c). The agreement is so good that the curves are indistinguishable. The same occurs for the entanglement entropy in Figure~\ref{Fig:FRM}d using $N_{pc} \sim \langle \exp[S_{vN} (t)] \rangle $. 

The entropies in Figure~\ref{Fig:FRM}c,d initially grow nearly quadratically, as can be verified by expanding Equation~(\ref{ShNpc}). This gives: 
\begin{equation}
S_{Sh,vN}^{GOE}(t \ll \sigma_{ini}^{-1}) \sim \left[ 1 - \ln \left( \frac{\sigma_{ini}^2}{N_{pc} } \right) - 2 \ln t \right] \sigma_{ini}^2 t^2 ,
\label{Eq:shortGOE}
\end{equation}
which matches the numerical results well for very short times. Subsequently, the data are very well fitted with a linear curve, as shown with circles in Figure~\ref{Fig:FRM}c,d. This makes a connection with the linear behavior, $S_{Sh,vN} \propto t$, studied theoretically in realistic systems~\cite{Eisert2010,Flambaum2001b,Berman2004,Santos2012PRL,Santos2012PRE,Haldar2016} and also observed experimentally~\cite{Kaufman2016}.

\subsection{Relaxation and Thermalization}\label{section2.4}

We say that an isolated finite quantum system equilibrated if, after a transient time, the initial state simply fluctuates around a steady state, remaining very close to it for most times (revivals may occur, but they take exceedingly long times to happen for large system sizes). Consider, for example, a few-body observable $O$ evolving according to the equation,
\begin{equation}
O(t) = \langle \Psi(0) | e^{i H t} O e^{-i H t} | \Psi(0) \rangle = \sum_{\alpha=1}^{ {\cal D} } |C_{ini}^{\alpha}|^2 O_{\alpha \alpha} + \sum_{\alpha \neq \beta=1}^{ {\cal D} } C_{ini}^{\beta *} C_{ini}^{\alpha} O_{\beta \alpha} e^{i (E_{\beta} - E_{\alpha}) t},
\label{Otime}
\end{equation}
where $O_{\beta \alpha} = \langle \psi_{\beta} | O | \psi_{\alpha} \rangle$ and $O_{\alpha \alpha} $ is the eigenstate expectation value of $O$. If the system equilibrates, there will be only fluctuations around the infinite time average $\overline{O}$. In the case of systems without degeneracies, as in GOE matrices, this average is given by the first term in Equation~(\ref{Otime}), 
\begin{equation}
\overline{O} = \sum_{\alpha=1}^{ {\cal D} } |C_{ini}^{\alpha}|^2 O_{\alpha \alpha} ,
\label{OtimeAve}
\end{equation}
with the second term leading to the fluctuations. 

The size of the fluctuations is crucial to determine whether the system indeed reaches equilibrium~\cite{Srednicki1996,Srednicki1999,Reimann2008,Short2011,Gramsch2012,Venuti2013,HeSantos2013,Zangara2013,KiendlARXIV}. For this, the fluctuations need to be small and decrease with ${\cal D}$. Small fluctuations certainly occur when the energy spacings are not zero and the products $C_{ini}^{\beta*} C_{ini}^{\alpha} $ and the off-diagonal elements $O_{\alpha \beta}$ are small. This is evidently the case for full random matrices, but also for realistic many-body quantum systems with interactions, where the fluctuations decrease exponentially with system size~\cite{Zangara2013}. We have found that the exponent of this decay depends on the level of delocalization of the initial state in the energy eigenbasis, with maximum exponents being found for full random matrices~\cite{Zangara2013}.

One can then talk about equilibration (or relaxation) without the need to invoke an environment. The loss of the initial coherence is, for all practical purposes, irreversible, because the recurrence time for systems with chaotic eigenstates is extremely long and increases with system size~\cite{Chirikov1986,Chirikov1997,Robinett2004}.

\subsubsection{Infinite-Time Averages}

The evolution of the Shannon entropy (\ref{SforFRM}) can be written in terms of the energy eigenbasis as:
\small{
\begin{eqnarray}
S_{Sh}(t) = &-& \sum_{k=1}^{\cal D} \left\{ \sum_{\alpha} \left| C_k^{\alpha} \right|^2 \left| C_{ini}^{\alpha} \right|^2
\ln \left[ 
\sum_{\gamma} \left| C_k^{\gamma} \right|^2 \left| C_{ini}^{\gamma} \right|^2 
+ \sum_{\gamma \neq \delta} C_k^{\gamma *} C_k^{\delta} C_{ini}^{\gamma} C_{ini}^{\delta *} e^{-i (E_{\gamma} - E_{\delta}) t}
\right] \right\} \nonumber \\
&-& \sum_{k=1}^{\cal D} \left\{ \sum_{\alpha \neq \beta} C_k^{\alpha *} C_k^{\beta} C_{ini}^{\alpha} C_{ini}^{\beta *} e^{-i (E_{\alpha} - E_{\beta}) t}
\ln \left[ 
\sum_{\gamma} \left| C_k^{\gamma} \right|^2 \left| C_{ini}^{\gamma} \right|^2 
+ \sum_{\gamma \neq \delta} C_k^{\gamma *} C_k^{\delta} C_{ini}^{\gamma} C_{ini}^{\delta *} e^{-i (E_{\gamma} - E_{\delta}) t}
\right] \right\} \nonumber .
\end{eqnarray}
}
Since on average $\sum_{\gamma} \left| C_k^{\gamma} \right|^2 \left| C_{ini}^{\gamma} \right|^2 
\gg \left| \sum_{\gamma \neq \delta} C_k^{\gamma *} C_k^{\delta} C_{ini}^{\gamma} C_{ini}^{\delta *} e^{-i (E_{\gamma} - E_{\delta}) t} \right|$, the infinite-time average of $S_{Sh}(t)$ is approximately:
{\begin{equation}
\overline{S}_{Sh} \sim - \sum_{k} \left[ \left(\sum_{\alpha} |C_{k}^{\alpha}|^2 |C_{ini}^{\alpha}|^2 \right) \ln 
\left(\sum_{\gamma} |C_{k}^{\gamma}|^2 |C_{ini}^{\gamma}|^2 \right) \right] .
\end{equation}

For GOE full random matrices, $\overline{S}_{Sh}$ is close to $S_{Sh}^{GOE}$ (\ref{ShGOE}). It is also similar to the diagonal entropy $S_d$, defined as~\cite{Polkovnikov2011,Santos2011PRL,Santos2012PRER},
\begin{equation}
S_d = -\sum_{\alpha} |C_{ini}^{\alpha}|^2 \ln |C_{ini}^{\alpha}|^2 .
\label{Sdiagonal}
\end{equation}
Equivalently, the infinite-time average of the entanglement entropy is close to $S_{vN}^{GOE}$ (\ref{SvGOE}).

The infinite-time average of the survival probability is the inverse of the participation ratio,
\begin{equation}
\overline{W}_{ini} = \sum_{\alpha} |C^{\alpha}_{ini} |^4  + \overline{\sum_{\alpha \neq \beta} |C^{\alpha}_{ini} |^2 |C^{\beta}_{ini} |^2 e^{i (E_{\beta} - E_{\alpha}) t} } \sim \frac{1}{PR_{ini}} .
\label{Eq:satF}
\end{equation}
For GOE full random matrices,
\begin{equation}
\overline{W}_{ini}^{GOE} \sim \frac{3}{ {\cal D}},
\label{Eq:satFgoe}
\end{equation}
which is reached in Figure~\ref{Fig:FRM}b. The size of the fluctuations decreases as $1/{\cal D}$. This is obtained from:
\begin{equation}
\sigma^2_{{W}_{ini} } = \overline{ |{W}_{ini} (t) - \overline{ {W}_{ini} (t) }|^2} =
\mathop{\sum_{\alpha \neq \beta} }_{\gamma \neq \delta} 
|C^{\alpha}_{\text{ini}}|^2 |C^{\beta}_{\text{ini}}|^2 |C^{\gamma}_{\text{ini}}|^2 |C^{\delta}_{\text{ini}}|^2
\overline{e^{i (E_{\alpha} - E_{\beta}+E_{\gamma} - E_{\delta}) t}} .
\end{equation}
On average, this term cancels out, except for $E_{\alpha} - E_{\beta} = E_{\delta} - E_{\gamma} $. Since there are no degeneracies of the energies or of the energy spacings in full random matrices, this equality requires that $E_{\alpha} = E_{\delta}$ and $E_{\beta} = E_{\gamma} $, which gives:
\begin{equation}
\sigma^2_{{W}_{ini} } = 
\sum_{\alpha} 
|C_{\alpha}^{\text{ini}}|^4 \sum_{\beta} |C_{\beta}^{\text{ini}}|^4 - \sum_{\alpha} 
|C_{\alpha}^{\text{ini}}|^8
\Longrightarrow \sigma_{{W}_{ini} } \sim \frac{1}{PR_{ini}}.
\end{equation}
The standard deviation of the temporal fluctuations of $W_{ini}$ coincides with $\overline{W}_{ini} $.

\subsubsection{Thermalization in Full Random Matrices}

After relaxation, the observable is said to have thermalized if its infinite-time average coincides with a thermodynamic average. Equality can only happen in the thermodynamic limit, but thermalization is already suggested if the two averages are close and further approach each other as the system size increases. The comparison between the two averages is made explicit with the~equation,
\begin{equation}
\overline{O} = \sum_{\alpha=1}^{ {\cal D} } |C_{ini}^{\alpha}|^2 O_{\alpha \alpha} \overbrace{=}^{? } O_{ME} \equiv \frac{1}{ {\cal{N}}_{E_{ini},\delta E} } \sum_{\substack{\alpha \\ |E_{ini}-E_\alpha|<\delta E}} \hspace{-0.5cm} O_{\alpha \alpha} 
\label{Eq:thermal}
\end{equation}
where $O_{ME}$
is the thermodynamic (microcanonical) average and ${\cal{N}}_{E_{ini},\delta E}$ is the number of energy eigenbases in the window $\delta E$ taken around $E_{ini}$. Equation~(\ref{Eq:thermal}) is valid when $O_{\alpha \alpha} $ for eigenstates close in energy agree with the microcanonical average, an idea that is at the heart of statistical mechanics and has recently become known as the eigenstate thermalization hypothesis (ETH) \cite{Deutsch1991,Srednicki1994,Rigol2008,rigol09STATa,rigol09STATb,Torres2013,He2013,Torres2014PRE,AlessioARXIV}. Notice that the ETH is not a condition for thermalization, but a statement of what one means by it.

Equation~(\ref{Eq:thermal}) is trivially satisfied for the GOE full random matrices. 
Since the eigenstates are random vectors, $O_{\alpha \alpha} $ is approximately the same for any eigenstate $|\psi_{\alpha}\rangle$, so it can be taken out of the sum and each $O_{\alpha \alpha} \sim O_{ME}$. The condition for thermalization is therefore quantum chaos, not only in the sense of level repulsion, but also in the sense of chaotic states~\cite{ZelevinskyRep1996,Santos2010PRE,RigolSantos2010,Santos2010PREb,Santos2011PRL,Santos2012PRL,Santos2012PRE,neuenhahn_marquardt_12,Torres2013,Borgonovi2016}. 

We stress that thermalization does not require an equal distribution of all probabilities, that is we do not need to have $|C_k^{\alpha}|^2 = 1/{\cal{D}}$, $S_{Sh}^{(\alpha)} = \ln{\cal D}$ and $S_{vN}^{(\alpha)} = \ln{\cal D}_A$ for all eigenstates, but the eigenstates that take part in the evolution of the initial state should be statistically close to these limits. We emphasize also that, contrary to full random matrices, not all eigenstates of real systems are chaotic, even when the Hamiltonian shows level repulsion. This is further discussed in Section \ref{section3}. Therefore, the  prerequisite for thermalization in real systems is to have equivalent values of $O_{\alpha \alpha}$ in the energy window sampled by the initial state. This is satisfied if the eigenstates in that window are very similar to random vectors, closely fulfilling Equations~(\ref{ShGOE}) and (\ref{SvGOE}). These two elements, energy window sampled by $|\Psi(0)\rangle$ and chaotic states, make evident the key role of the interplay between initial state and Hamiltonian in the analysis of thermalization~\cite{Torres2013,He2013}. The third important element is the observable, as discussed also in~\cite{Jensen1985}. The observables considered in the studies of thermalization in realistic systems are few-body observables.

As Dyson had already anticipated, the development of random matrix theory marked the beginning of a ``new kind of statistical mechanics'' \cite{Dyson1962b}, but at those early stages, the link between equilibrium and nonequilibrium dynamics, as put forward by Equation~(\ref{Eq:thermal}), was not yet well established. Current studies analyze Equation~(\ref{Eq:thermal}) for specific few-body observables, take into account the properties of the initial state and consider realistic systems that may not even be disordered, but in which strong interactions lead to stochastic behavior.

In the cases of the particular quantities studied here, quantum chaos guarantees that in realistic systems, $\overline{S}_{Sh} \sim \ln (a {\cal D})$ and $\overline{S}_{vN} \sim \ln (a {\cal D}_A)$, with $a$ being a constant. The fact that these entropies saturate
to values that follow a volume law imply that they approach thermodynamic entropies as the system size increases~\cite{Santos2011PRL,Santos2012PRER}. For chaotic initial states, we also have $\overline{W}_{ini} \sim {\cal D}^{-1}$, which assures thermalization.


\section{Realistic Integrable and Chaotic Models}\label{section3}

Full random matrices do not model realistic systems. The Hamiltonian matrices describing realistic systems are usually sparse and banded, due to the presence of few-body (often~only~two-body) interactions, and in various cases, they do not even have random elements. Attempts to bring random matrix theory closer to realistic systems started already with Wigner, with the introduction of the so-called Wigner banded random matrices~\cite{Wigner1955}. In these matrices, only the elements within a bandwidth around the diagonal are nonzero. Other similar approaches include two-body random ensembles and power law banded random matrices~\cite{French1970,Bohigas1971,Brody1981,Mirlin1996}.

To bring the discussion of Section~\ref{section2} down to earth, we consider a one-dimensional spin-1/2 model, which is similar to the systems studied experimentally with cold atoms, ion traps and nuclear magnetic resonance platforms. It is described by the Hamiltonian:
\begin{equation}
H = \epsilon_1 J S_{1}^z + d J S_{\lfloor L/2 \rfloor }^z + H_{\text{XXZ}}+ \lambda H_{\text{NNN}},
\label{ham}
\end{equation}
where:
\begin{eqnarray}
&& H_{\text{XXZ}} = J\sum_{n=1}^{L-1} \left( 
S_n^x S_{n+1}^x + S_n^y S_{n+1}^y +
\Delta S_n^z S_{n+1}^z \right), 
 \\
&& H_{\text{NNN}} = J\sum_{n=1}^{L-2} \left(S_n^x S_{n+2}^x + S_n^y S_{n+2}^y
+ \Delta S_n^z S_{n+2}^z \right). 
\end{eqnarray}
Above, $\hbar=1$, $S^{x,y,z}_n$ are the spin operators on site $n$; $L$ is the total even number of sites in the chain; $J$ is the coupling strength; and $\Delta$ is the anisotropy parameter. The Hamiltonian contains nearest-neighbor (NN) couplings and next-nearest-neighbor (NNN) couplings if $\lambda \neq 0$. We always include a small defect $\epsilon_1=0.1$ (Zeeman splitting 0.1 larger than that of the other sites) on the first site to break parity and spin reversal symmetries. A defect of amplitude $d$ may also be included in the middle of the chain, on site $n=\lfloor L/2 \rfloor$. We consider open boundary conditions. The total spin in the $z$-direction, ${\cal S}^z$, is conserved. We study the largest subspace, ${\cal S}^z=0$, so ${\cal D}=L!/(L/2)!^2$. In all of the figures below, except for Figure~\ref{Fig:power law}, we consider $L=16$, so ${\cal D} =12870$ and ${\cal D}_A=256$. This explains the choices of ${\cal D}$ and ${\cal D}_A$ made for the figures for full random matrices in Section~\ref{section2}.

All of the parameters are taken as positive, and $J=1$ sets the energy scale. When $\lambda, d=0$ and $\Delta \neq 0$ (we fix $\Delta =0.48$), the Hamiltonian is integrable and known as the XXZ model. [XXZ refers to models where the coupling strengths in the $x$ and $y$ directions are the same and different from that in the $z$ direction. The isotropic version is known as XXX.] The inclusion of NNN couplings or of a defect away from the borders of the chain breaks the integrability of the XXZ Hamiltonian~\cite{Santos2004,Torres2014PRE}. Level repulsion occurs for the parameters considered here: $\lambda=1$ ($d=0$) and $d =0.9$ ($\lambda=0$). We refer to these two cases as the NNN model and the defect model, respectively. In common with the XXZ Hamiltonian, the first is clean, and the second has only NN couplings.

For any value of the parameters in $H$ (\ref{ham}), the density of states is Gaussian, as is typical of many-body quantum systems with two-body interactions. The Gaussian shape is a consequence of combinatorics and the central limit theorem. This is the first crucial difference with respect to full random matrices that is worth emphasizing. The distributions are shown in Figure~\ref{Fig:DOS}a--c for the three considered spin models. The Gaussian energy distribution is symmetric when $\epsilon_1,d,\Delta,\lambda=0$, while defects, anisotropy and $\lambda$ make it asymmetric.

\begin{figure}[ht]
\centering
\includegraphics*[width=8.3cm]{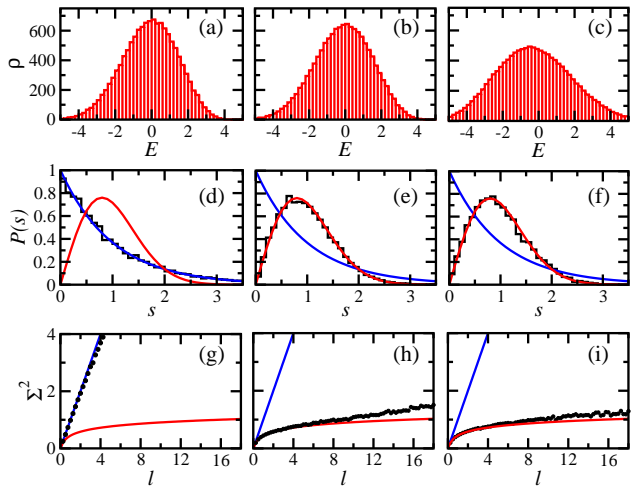}
\caption{Top panels: density of states for the XXZ 
 (\textbf{a}), defect (\textbf{b}) and next-nearest-neighbor (NNN) (\textbf{c}) models. Middle and bottom panels: level spacing distribution (\textbf{d},\textbf{e},\textbf{f}) and level number variance (\textbf{g},\textbf{h},\textbf{i}), respectively, for the same models in (a,b,c).
Open boundaries, $\epsilon_1=0.1$, $d=0.9$, $\Delta=0.48$, $\lambda=1$, $L=16$, ${\cal S}^z=0$.}
\label{Fig:DOS}
\end{figure} 

Level spacing distribution and level number variance are shown in the middle and bottom panels of Figure~\ref{Fig:DOS}, clearly distinguishing the integrable model from the chaotic ones. The XXZ model presents additional symmetries when the anisotropy reaches the roots of unity, $\Delta= \cos (\pi/\mu)$, where $\mu \geq 2$ is an integer. The largest amount of degeneracies occur for $\mu=2$, but significant amounts exist also for $\mu=3$ \cite{Zangara2013}. This explains why we chose $\Delta=0.48$ instead of 1/2, even though the presence of the small defect at the border, $\epsilon_1$, also prevents too many degeneracies.

The initial impression is that there is not much difference between Figure~\ref{Fig:DOS}e,f and Figure~\ref{Fig:FRMstatic}a. However, visibly, the spectrum of the GOE full random matrix is more rigid than those for the chaotic spin models, as evident by comparing Figure~\ref{Fig:FRMstatic}b with Figure~\ref{Fig:DOS}h,i. This reinforces the importance of considering different signatures of quantum chaos when comparing models.


\subsection{Delocalization and Entanglement Measures: Basis Dependence}

The Gaussian density of states implies that there are more states in the middle of the spectrum than at the edges. The states away from the borders are then expected to be more delocalized. This energy dependence on the level of delocalization of $|\psi_{\alpha}\rangle $ is another crucial difference between realistic systems and full random matrices. 

To compute $S_{Sh}^{(\alpha)}$, we need to choose a basis representation. This choice is made according to the problem we are studying. If one is interested in localization in real space, the site basis (known in quantum information as computational basis) is the natural choice. It corresponds to states where on each site, the spin points either up or down in the $z$-direction. However, if the interest is in the level of chaoticity of the states, the mean-field basis is the most appropriate one. The mean-field basis is defined by the integrable (regular) part of the Hamiltonian. In the case of the NNN and defect models, a good choice for the mean-field basis is to use the eigenstates of the XXZ model. We also verify that the differences obtained when we consider the eigenstates of the XX model are not significant. For the XXZ model, we use the eigenstates of the XX model~\cite{Santos2012PRL,Santos2012PRE} as a way to analyze how the level of complexity of $|\psi_{\alpha}\rangle $ increases with the Ising interaction.

In Figure~\ref{Fig:ShSv}a--c, we show the Shannon entropy for the XXZ, defect and NNN models in the site and mean-field basis. Independent of the basis, larger fluctuations are observed for the integrable model~\cite{Santos2010PRE, RigolSantos2010,Santos2010PREb}, where the eigenstates are further from random vectors than in chaotic models.
The values of $S_{Sh}^{(\alpha)}$ for the XXZ Hamiltonian are also smaller overall than those for the chaotic models. 

\begin{figure}[ht]
\centering
\includegraphics*[width=11cm]{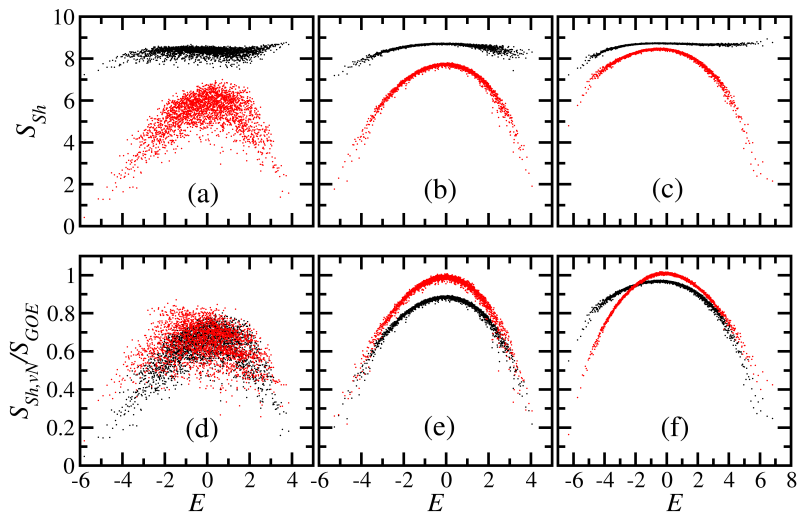}
\caption{Top panels: Shannon entropy in the site basis (black symbols) and in the mean-field basis (red symbols) for the XXZ (\textbf{a}), defect (\textbf{b}) and NNN (\textbf{c}) models. Bottom panels: normalized entanglement entropy (black symbols) and normalized Shannon entropy in the mean-field basis (red symbols) for the same models as in (a,b,c). Parameters for all panels are the same as in Figure~\ref{Fig:DOS}.}
\label{Fig:ShSv}
\end{figure} 

In the mean-field basis, one clearly sees that the eigenstates of the chaotic models close to the middle of the spectrum are very complex, with $S_{Sh}^{(\alpha)}$ approaching the full random matrix value $S_{Sh}^{GOE} $~(\ref{ShGOE}). As we move towards the edges of the spectrum, the states become more localized. This is why thermalization is not expected to take place close to the borders of the spectrum~\cite{Santos2010PRE, RigolSantos2010,Santos2010PREb,Torres2013}. We note that the somewhat large values of $S_{Sh}^{(\alpha)}$ obtained for the NNN model at low energies occurs only when $\lambda$ is large, but they are small when $\lambda \sim 0.5$ \cite{Santos2012PRE,Torres2014PRE}. This may suggest that in that region of the spectrum and for $\lambda \rightarrow 1$, the basis considered is not the best mean-field~basis.

In the site basis, the energy dependence is not so obvious, and $S_{Sh}^{(\alpha)}$ even surpasses $S_{Sh}^{GOE} $. Values above $S_{Sh}^{GOE} $ indicate that the chosen basis is not a good mean-field basis. For the models considered, the site basis is therefore not the correct one to evaluate the proximity of the eigenstates to chaotic~vectors. 

In Figure~\ref{Fig:ShSv}d--f, we compare the normalized entanglement entropy $S_{vN}^{(\alpha)}/S_{vN}^{GOE} $ with the normalized Shannon entropy $S_{Sh}^{(\alpha)}/S_{Sh}^{GOE}$ for the same three models. The behavior of the two entropies is quite similar, so either one could be used in the analysis of the complexity of the states. Each entropy has its own advantages and disadvantages. In contrast to $S_{vN}^{(\alpha)}$, the basis needs to be carefully chosen when computing $S_{Sh}^{(\alpha)}$, but the latter is computationally less expensive, since no partial trace needs to be performed.


\subsection{Dynamics at Intermediate Times: Generic Behaviors}

The dynamics of the spin model is initiated after its preparation in an eigenstate of a certain Hamiltonian $H_{ini}$. The state is then left to evolve according to $H$ (\ref{ham}), where $H= H_{ini} + \nu V $ and $\nu$ is the perturbation strength.

As discussed in Section~\ref{section2.3.1}, the behavior of the survival probability is determined by the LDOS, $P_{ini,ini}(E)$. As $\nu$ increases from zero, the LDOS broadens~\cite{Torres2014NJP,Torres2014PRE,TorresProceed}. When the couplings (elements of~$\nu V$) become larger than the mean level spacing, the envelope of the LDOS acquires a Lorentzian shape, and its Fourier transform leads to the exponential decay of $W_{ini}(t)$. By further increasing $\nu$, the LDOS eventually becomes Gaussian, causing the Gaussian decay of $W_{ini}(t)$ \cite{Torres2014PRA,Torres2014NJP,Torres2014PRE,TorresProceed,Torres2014PRAb,TorresKollmar2016,Torres2015,Torres2016BJP,TavoraARXIV}. This is not the mere quadratic decay observed for very short times for any perturbation strength, which emerges from the expansion:
\begin{equation}
W_{ini}(t<<\sigma_{ini}^{-1}) \sim \left| \langle \Psi(0) | 1 -i H t | \Psi(0) \rangle \right|^2 \sim 1- \sigma_{ini}^2 t^2,
\end{equation}
but a true Gaussian decay holding for larger times (see also \cite{Izrailev2006} and the references therein). The Gaussian LDOS reflects the Gaussian density of states and corresponds to the maximum spread in energy of the initial state. This limiting scenario is reached by strongly perturbing the many-body quantum system, and it does not matter whether or not there is level repulsion.

Several examples of Gaussian LDOS and Gaussian decay of $W_{ini}(t)$ have been shown for the clean XXZ and NNN models in~\cite{Torres2014PRA,Torres2014NJP,Torres2014PRE,TorresProceed,Torres2014PRAb,TorresKollmar2016} and also for disordered models~\cite{Torres2015,Torres2016BJP}. In Figure~\ref{Fig:GaussDefect}, we compare the integrable XXZ model with the chaotic defect model. As the initial state, $|\Psi(0)\rangle = |\phi_{ini} \rangle$, we consider the N\'eel state, $|\downarrow \uparrow \downarrow \uparrow \downarrow \uparrow \downarrow \uparrow\ldots \rangle$, where the $z$-polarization of the spins alternates from one site to the other. This state is often prepared in experiments with cold atoms. The evolution under both models is equivalent to having had a very strong perturbation, where $\Delta$ is changed from $\infty$ to~$0.48$. 

The envelopes of the LDOS in Figure~\ref{Fig:GaussDefect}a,c agree very well with a Gaussian of width: 
\begin{equation}
\sigma_{ini} = \sqrt{\sum_{k \neq ini } |\langle \phi_k | H | \phi_{ini} \rangle |^2 } = \frac{J}{2} \sqrt{L-1},
\label{sigmaNS}
\end{equation}
 centered at: 
 \begin{equation}
 E_{ini} = \langle \phi_{ini} | H | \phi_{ini}\rangle = \frac{\epsilon_1}{2} + (-1)^{\frac{L}{2}} \frac{d}{2}-\frac{J \Delta}{4} (L-1) ,
 \label{ENS}
 \end{equation}
with $d$ being zero for the XXZ model and 0.9 for the defect model. For both models, the chosen initial state is further from the middle of the spectrum and, therefore, less filled than for the NNN model,~where: 
\begin{equation}
E_{ini}=\frac{\epsilon_1}{2} + \frac{J \Delta}{4} [- (L-1) + (L-2) \lambda].
\label{EiniNNN}
\end{equation}
Comparing Figure~\ref{Fig:GaussDefect}a and \ref{Fig:GaussDefect}c, we also see that the LDOS is better filled for the defect model than for the XXZ model, since the first is chaotic and has $ E_{ini}$ slightly closer to the middle of the spectrum. We also note that the width of the initial state is sub-extensive, $\sigma_{ini} \propto \sqrt{L}$, so the initial state is narrow in energy. This is a necessary condition for thermalization in real systems~\cite{Srednicki1994,Rigol2008}.

In Figure~\ref{Fig:GaussDefect}b,d, we show the evolution of the survival probability. The numerical results are very close to the analytical Gaussian decay, $W_{ini}(t) = \exp( - \sigma_{ini}^2 t^2)$, until the curves cross the saturation line $1/PR_{ini}$ (\ref{Eq:satF}). In this limit of strong perturbation and initial states with energy $E_{ini}$ away from the edges of the spectrum, the behavior of the survival probability is general and equivalent for both integrable and chaotic many-body quantum models. For the intermediate time scales considered in Figure~\ref{Fig:GaussDefect}, level repulsion does not play any important role.

\begin{figure}[ht]
\centering
\includegraphics*[width=10cm]{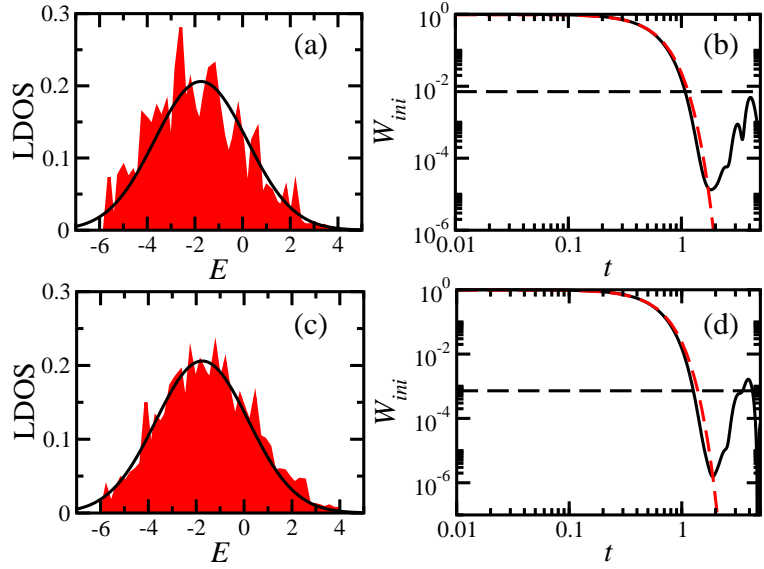}
\caption{N\'eel state under the XXZ model (\textbf{a},\textbf{b}), and defect model (\textbf{c},\textbf{d}). (a,c) Numerical results for the LDOS (shaded area) and Gaussian envelope (solid line) with $\sigma_{ini}$ from Equation~(\ref{sigmaNS}) and $E_{ini}$ from Equation~(\ref{ENS}). (b,d) Numerical results for the survival probability (solid line), analytical expression $W_{ini}(t) = \exp( - \sigma_{ini}^2 t^2)$ (dashed line) with $\sigma_{ini}$ from Equation~(\ref{sigmaNS}) and saturation value (horizontal line) given by $1/PR_{ini}=\sum_{\alpha} |C_{ini}^{\alpha}|^4$ [Equation~(\ref{Eq:satF})].
Parameters as in Figure~\ref{Fig:DOS}.}
\label{Fig:GaussDefect}
\end{figure}

In Figure~\ref{Fig:EntropiesLinear}, we compare the evolution of the Shannon entropy (top panels) and the entanglement entropy (bottom panels) for the N\'eel state under the XXZ (a,d), defect (b,e) and NNN (c,f) models. At very short times, $t \ll \sigma_{ini}^{-1}$, the entropies increase nearly quadratically~\cite{Torres2014NJP}. Later, as in full random matrices, the entropy growth becomes linear in time and remains as such until close to saturation. This linear behavior is expected for initial states far from the edges of the spectrum and seems independent of the presence or absence of level repulsion. What one observes, however, is a dependence on the energy of the initial state~\cite{Torres2014PRAb} and on the connectivity of the Hamiltonian~\cite{Torres2014NJP}. For the NNN model, $\sigma_{ini} $ is still the same as in Equation~(\ref{sigmaNS}), but since $E_{ini}$ from Equation~(\ref{EiniNNN}) is closer to the middle of the spectrum, where there are more eigenstates and they are more delocalized than those from the XXZ and defect model (cf. Figure~\ref{Fig:ShSv}), the slope of the linear entropy increase is larger in Figure~\ref{Fig:EntropiesLinear}c,f than in Figure~\ref{Fig:EntropiesLinear}a,b,d,e.

The saturation values of the entropies for the GOE full random matrices are indicated in Figure~\ref{Fig:EntropiesLinear} with horizontal dashed lines. For the XXZ model, the entropies saturate to values smaller than $S_{Sh,vN}^{GOE}$. For the defect model, $\overline{S}_{Sh}$ surpasses $S_{Sh}^{GOE}$. For the NNN model, both $\overline{S}_{Sh}$ and $\overline{S}_{vN}$ are above the results for full random matrices. This is in contrast with previous studies for the evolutions of the Shannon entropy under the NNN model that start with mean-field basis vectors and analyze the spreading over other mean-field basis~\cite{Santos2012PRL,Santos2012PRE}. In these cases, the saturation values are not larger than $S_{Sh,vN}^{GOE}$, but very close to it. 

Another difference between the results for site basis initial states and mean-field initial states is the agreement with Equation~(\ref{ShNpc}), which occurs for the latter~\cite{Santos2012PRL,Santos2012PRE}, but not for the former~\cite{Torres2014NJP}. The approximation in Equation~(\ref{ShNpc}) seems appropriate for initial states that are directly coupled with many other basis vectors, that is their connectivity is $\propto {\cal D}$. This is not the case of site basis vectors. The N\'eel state, for instance, is directly coupled with only $L-1$ states.

\begin{figure}[ht]
\centering
\includegraphics*[width=11cm]{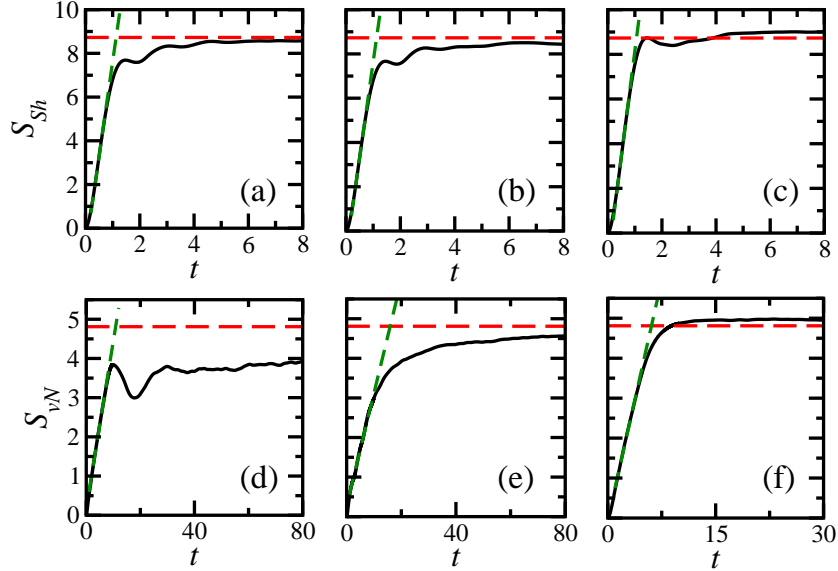}
\caption{Evolution of the Shannon entropy (top panels) and entanglement entropy (bottom panels) for the N\'eel state under the XXZ (\textbf{a},\textbf{d}), defect (\textbf{b},\textbf{e}) and NNN (\textbf{c},\textbf{f}) models. Numerical results (solid lines) and fitted linear growth $S_{Sh,vN} = a_{Sh,vN} +b_{Sh,vN} t$ (green dashed lines) with $a_{Sh}= -1.04,-1.04, -1.31$ and $b_{Sh}=8.67, 8.67, 9.47$ from (a--c) and $a_{vN}=0.17,0.21,-0.06$ and $b_{vN}=0.43, 0.30, 0.80$ from (d--f). Horizontal dashed lines indicate $S_{Sh}^{GOE} \sim \ln(0.48 {\cal D})$ (top panels) and $S_{vN}^{GOE} \sim \ln(0.48 {\cal D}_A)$ (bottom panels). Parameters as in Figure~\ref{Fig:DOS}.}
\label{Fig:EntropiesLinear}
\end{figure}


\subsection{Dynamics at Long Times}

At long times, the evolution of $W_{ini}(t)$ slows down. The decay of the survival probability becomes necessarily a power law, $W_{ini}(t\gg \sigma_{ini}^{-1}) \propto t^{-\gamma}$. It is the tails of the LDOS that determine the value of the power law exponent $\gamma$. Good filling implies that, despite the discreteness of the spectrum, the LDOS can be treated as a nearly continuous function. The behavior of $W_{ini}(t)$ at long times can then be obtained from the Fourier transform of the Gaussian $P_{ini,ini}(E)$, but taking into account also the unavoidable energy bounds $E_{low}$ and $E_{up}$ of the spectrum~\cite{TavoraARXIV}, 
\begin{equation}
W_{ini}(t) = \frac{1}{\sqrt{2 \pi \sigma_{ini}^2 }} \int_{E_{low}}^{E_{up}} e^{- (E-E_{ini})^2/(2 \sigma_{ini}^2)} e^{-i E t} dE \Longrightarrow W_{ini}(t \gg \sigma_{ini}^{-1}) \propto t^{-2}.
\end{equation}

The above power law decay with exponent $\gamma=2$ can be seen in Figure~\ref{Fig:power law}a and also in other chaotic initial states studied in~\cite{TavoraARXIV}. The exponent is smaller than that for full random matrices, where $\gamma=3$ (see Equation~(\ref{Eq:powerGOE})), but larger than what we obtain for systems undergoing localization, where the LDOS is very sparse and $\gamma<1$~\cite{Torres2015,Torres2016BJP,TavoraARXIV}. In many-body quantum systems, exponents $\gamma \geq 2$ indicate that the LDOS is very well filled and that the components $C_{ini}^{\alpha}$ are close to random numbers from a Gaussian distribution. In this scenario, we should expect thermalization to occur~\cite{TavoraARXIV}.

In Figure~\ref{Fig:power law}b, we show the evolution of the Shannon entropy. The symbols represent a fitted linear curve. The log-log plot makes it very clear that the linear growth holds only for times that are not too short. The dot-dashed line corresponds to Equation~(\ref{Eq:shortGOE}). It is parallel to the numerical data nearly up to the point where the linear behavior develops. For the curve of Equation~(\ref{Eq:shortGOE}) to actually coincide with the numerical data, we would need to substitute $N_{pc}$ by a small value, $\sim 20$. This happens when the initial state is a site basis vector, not when it is a mean-field basis vector.

In analogy with the onset of the power law decay for $W_{ini}(t)$, we might wonder whether at long times, Figure~\ref{Fig:power law}b actually detects a behavior different from the linear growth. Right before saturation, between $t\sim1$ and $t\sim 3$, a small oscillation is visible. This is also noticeable in Figure~\ref{Fig:EntropiesLinear}, especially in Panel (c). This time interval coincides with that for the survival collapse observed for the survival probability. The collapse~\cite{TavoraARXIV,Fiori2006,Fiori2009} corresponds to an interference between the Gaussian decay and the emergence of the power law decay, which pushes $W_{ini}(t)$ very much below the saturation line (see Figure~\ref{Fig:power law}a and also Figure~\ref{Fig:GaussDefect}b,d). Thus, even though the small oscillation requires further studies, it may be an indication of an interference between two different behaviors, as in the case of the survival~collapse.

\begin{figure}[ht]
\centering
\includegraphics*[width=12cm]{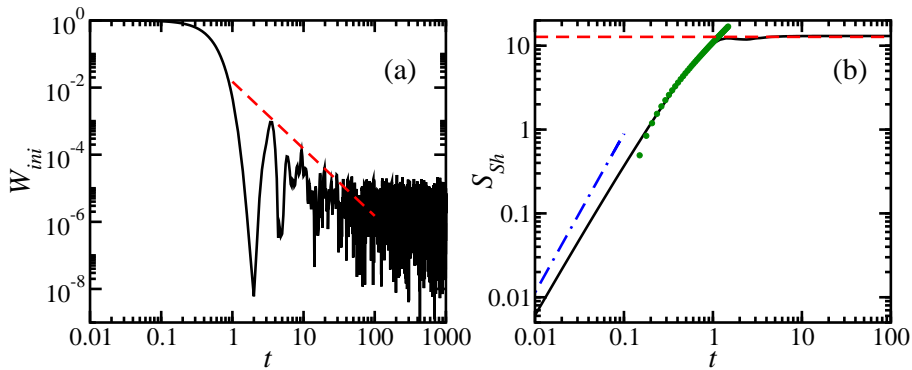}
\caption{Survival probability (\textbf{a}) and evolution of the Shannon entropy (\textbf{b}) for the N\'eel state under the NNN model with the parameters of Figure~\ref{Fig:DOS}, but $L=22$. In (a), the numerical result is given by the solid line and the power law decay $\propto t^{-2}$ by the dashed line. In (b), the numerical result is given by the solid line; Equation~(\ref{Eq:shortGOE}) by a dot-dashed line; and the linear increase, $S_{Sh} \propto t$, by the symbols.}
\label{Fig:power law}
\end{figure}

\section{Discussion}

We compared the static and dynamical properties of GOE full random matrices and finite isolated many-body quantum systems described by one-dimensional spin-1/2 models with two-body interactions. This comparison is useful, because for full random matrices, analytical expressions can be derived and then used as references and bounds for the analysis of realistic models.
Our main findings are itemized below:

\begin{enumerate}
\item The results for the von Neumann entanglement entropy $S_{vN}$, which is a concept employed in quantum information science, and for the Shannon information entropy $S_{Sh}$, which is generally used as a measurement of the degree of delocalization of quantum states, were very similar. Thus, either one can be used to measure the level of complexity of the eigenstates. The advantage of the Shannon entropy is that it is computationally less expensive than the entanglement entropy. The disadvantage is that it is strongly dependent on the basis chosen.
\vspace{6pt}
\item For full random matrices, all eigenstates are pseudo-random vectors and therefore lead to the same values of $S_{Sh}^{(\alpha)}$, but the results for realistic systems depend on the region of the spectrum and on the basis selected. 
 \vspace{6pt}
\item An analytical expression was given for full random matrices for the time evolution of both entropies. It agrees extremely well with the numerical results. For the spin systems; this expression gives an upper bound for $S_{Sh} (t)$ and $S_{vN}(t)$.
\vspace{6pt}
\item At short times, $S_{Sh} (t)$ and $S_{vN} (t)$ show a nearly quadratic behavior. It is only at longer times that the linear increase, $S_{Sh,vN} (t) \propto t$, develops. These two behaviors seem to be independent of the presence or absence of level repulsion.
\end{enumerate}

We also reviewed some of our previous studies and discussions. They include:

\begin{enumerate}
\item In realistic chaotic models, the spectrum is not as rigid as that of full random matrices. When comparing different chaotic models, it is appropriate to compare different signatures of chaos, such as those that detect short-range and also long-range correlations.
\vspace{6pt}
\item Analytical expressions for the decay of the survival probability, $W_{ini}(t)$, were given for full random matrices and for the spin systems. For realistic models, integrable or chaotic, the decay at short times is Gaussian when the perturbation that takes the system out of equilibrium is strong. The decay is faster for full random matrices.
\vspace{6pt}
\item At long times, the decay of the survival probability becomes a power law, $W_{ini}(t) \propto t^{-\gamma}$, with $\gamma=3$ for full random matrices and $\gamma^{max}=2$ for the spin systems. The emergence of a power law decay at long times should have interesting consequences for problems associated with quantum information science and foundations of quantum mechanics. One should expect, for example, that external actions on the system, such as measurements, performed at long times may change the power law decay and recover Gaussian or exponential decays. This idea was explored in~\cite{Lawrence2002} for a one-body system interacting with an environment. It would be worth extending it also to many-body quantum systems.
\vspace{6pt}
\item Equilibration and thermalization are trivially reached under full random matrices. In realistic models, the absence of degeneracies and the presence of chaotic states in the energy window sampled by the initial state are both key elements for achieving thermal equilibrium.
\end{enumerate}

\section{Materials and Methods}

The numerical methods used include exact diagonalization and Expokit~\cite{Sidje1998,Expokit}. Exact diagonalization was used for Hamiltonian matrices with ${\cal D}<$ 20,000, and Expokit was employed for the dynamics of systems with larger ${\cal D}$. 

\vspace{6pt}

\acknowledgments{EJTH acknowledges funding from CONACyT, PRODEP-SEP and Proyectos VIEP-BUAP 2016, Mexico. EJTH is also grateful to LNS-BUAP for allowing use of their supercomputing facility. This work was supported by the NSF grant No.~DMR-1147430.}

\renewcommand\bibname{References}


\begin{thebibliography}{999}

\providecommand{\natexlab}[1]{#1}

\bibitem[Preskill(2013)]{PreskillProceed}
Preskill, J.
\newblock Quantum computing and the entanglement frontier.
\newblock In {\em The Theory of the Quantum Worlds,} Gross,~D., Henneaux, M., Sevrin,
  A., Eds.; World Scientific: Singapore, 2013; pp. 63--90.

\bibitem[White(1992)]{White1992}
White, S.R.
\newblock Density matrix formulation for quantum renormalization groups.
\newblock {\em Phys. Rev. Lett.} {\bf 1992}, {\em 69},~2863.

\bibitem[Feiguin and White(2005)]{Feiguin2005a}
Feiguin, A.E.; White, S.R.
\newblock Time-step targeting methods for real-time dynamics using the density
  matrix renormalization group.
\newblock {\em Phys. Rev. B} {\bf 2005}, {\em 72},~020404.

\bibitem[Eisert \em{et~al.}(2010)Eisert, Cramer, and Plenio]{Eisert2010}
Eisert, J.; Cramer, M.; Plenio, M.B.
\newblock \textit{Colloquium}: Area laws for the entanglement entropy.
\newblock {\em Rev. Mod. Phys.} {\bf 2010}, {\em 82},~277.

\bibitem[Zelevinsky \em{et~al.}(1996)Zelevinsky, Brown, Frazier, and
  Horoi]{ZelevinskyRep1996}
Zelevinsky, V.; Brown, B.A.; Frazier, N.; Horoi, M.
\newblock The nuclear shell model as a testing ground for many-body quantum
  chaos.
\newblock {\em Phys. Rep.} {\bf 1996}, {\em 276},~85--176.

\bibitem[Borgonovi \em{et~al.}(2016)Borgonovi, Izrailev, Santos, and
  Zelevinsky]{Borgonovi2016}
Borgonovi, F.; Izrailev, F.M.; Santos, L.F.; Zelevinsky, V.G.
\newblock Quantum chaos and thermalization in isolated systems of interacting
  particles.
\newblock {\em Phys. Rep.} {\bf 2016}, {\em 626},~1--58.

\bibitem[D'Alessio \em{et~al.}()D'Alessio, Kafri, Polkovnikov, and
  Rigol]{AlessioARXIV}
D'Alessio, L.; Kafri, Y.; Polkovnikov, A.; Rigol, M.
\newblock From quantum chaos and eigenstate thermalization to statistical
  mechanics and thermodynamics.
\newblock {\bf 2015}, arXiv:1509.06411.

\bibitem[Guhr \em{et~al.}(1998)Guhr, Mueller-Gr\"oeling, and
  Weidenm\"uller]{Guhr1998}
Guhr, T.; Mueller-Gr\"oeling, A.; Weidenm\"uller, H.A.
\newblock Random-matrix theories in quantum physics: Common concepts.
\newblock {\em Phys. Rep.} {\bf 1998}, {\em 299},~189--425.

\bibitem[Gubin and Santos(2012)]{Gubin2012}
Gubin, A.; Santos, L.F.
\newblock Quantum chaos: An introduction via chains of interacting spins 1/2.
\newblock {\em Am. J. Phys.} {\bf 2012}, {\em 80},~246--251.

\bibitem[Santos \em{et~al.}(2004)Santos, Rigolin, and
  Escobar]{SantosEscobar2004}
Santos, L.F.; Rigolin, G.; Escobar, C.O.
\newblock Entanglement versus chaos in disordered spin systems.
\newblock {\em Phys. Rev.~A} {\bf 2004}, {\em 69},~042304.

\bibitem[Mejia-Monasterio \em{et~al.}(2005)Mejia-Monasterio, Benenti, Carlo,
  and Casati]{Monasterio2005b}
Mejia-Monasterio, C.; Benenti, G.; Carlo, G.G.; Casati, G.
\newblock Entanglement across a transition to quantum chaos.
\newblock {\em Phys. Rev. A} {\bf 2005}, {\em 71},~062324.

\bibitem[Lakshminarayan and Subrahmanyam(2005)]{Lakshminarayan2005}
Lakshminarayan, A.; Subrahmanyam, V.
\newblock Multipartite entanglement in a one-dimensional time-dependent Ising
  model.
\newblock {\em Phys. Rev. A} {\bf 2005}, {\em 71},~062334.

\bibitem[Brown \em{et~al.}(2008)Brown, Santos, Starling, and Viola]{Brown2008}
Brown, W.G.; Santos, L.F.; Starling, D.J.; Viola, L.
\newblock Quantum chaos, localization, and entanglement in disordered
  {H}eisenberg Models.
\newblock {\em Phys. Rev. E} {\bf 2008}, {\em 77},~021106.

\bibitem[Dukesz \em{et~al.}(2009)Dukesz, Zilbergerts, and Santos]{Dukesz2009}
Dukesz, F.; Zilbergerts, M.; Santos, L.F.
\newblock Interplay between interaction and (un)correlated disorder in
  one-dimensional many-particle systems: delocalization and global
  entanglement. {\em New J. Phys.} {\bf 2009}, {\em 11},~043026.

\bibitem[Giraud \em{et~al.}(2009)Giraud, Martin, and Georgeot]{Giraud2009}
Giraud, O.; Martin, J.; Georgeot, B.
\newblock Entropy of entanglement and multifractal exponents for random states.
\newblock {\em Phys. Rev. A} {\bf 2009}, {\em 79},~032308.

\bibitem[Izrailev and Casta{\~{n}}eda-Mendoza(2006)]{Izrailev2006}
Izrailev, F.M.; Casta{\~{n}}eda-Mendoza, A.
\newblock Return probability: Exponential versus Gaussian decay.
\newblock {\em Phys. Lett.~A} {\bf 2006}, {\em 350},~355--362.

\bibitem[Torres-Herrera and Santos(2014)]{Torres2014PRA}
Torres-Herrera, E.J.; Santos, L.F.
\newblock Quench dynamics of isolated many-body quantum systems.
\newblock {\em Phys. Rev. A} {\bf 2014}, {\em 89},~043620.

\bibitem[Torres-Herrera \em{et~al.}(2014)Torres-Herrera, Vyas, and
  Santos]{Torres2014NJP}
Torres-Herrera, E.J.; Vyas, M.; Santos, L.F.
\newblock General features of the relaxation dynamics of interacting quantum
  systems.
\newblock {\em New J. Phys.} {\bf 2014}, {\em 16},~063010.

\bibitem[Torres-Herrera and Santos(2014{\natexlab{a}})]{TorresProceed}
Torres-Herrera, E.J.; Santos, L.F.
\newblock Isolated many-body quantum systems far from equilibrium: Relaxation
  process and thermalization.
In Proceedings of the Fourth Conference on Nuclei and Mesoscopic Physics, East Lansing, MI, USA, 5–9 May 2014.

\bibitem[Torres-Herrera and Santos(2014{\natexlab{b}})]{Torres2014PRAb}
Torres-Herrera, E.J.; Santos, L.F.
\newblock Nonexponential fidelity decay in isolated interacting quantum
  systems.
\newblock {\em Phys. Rev. A} {\bf 2014}, {\em 90},~033623.

\bibitem[Torres-Herrera and Santos(2014{\natexlab{c}})]{Torres2014PRE}
Torres-Herrera, E.J.; Santos, L.F.
\newblock Local quenches with global effects in interacting quantum systems.
\newblock {\em Phys.~Rev. E} {\bf 2014}, {\em 89},~062110.

\bibitem[Torres-Herrera \em{et~al.}(2016)Torres-Herrera, Kollmar, and
  Santos]{TorresKollmar2016}
Torres-Herrera, E.J.; Kollmar, D.; Santos, L.F.
\newblock Relaxation and thermalization of isolated many-body quantum systems.
\newblock {\em Phys. Scr. } {\bf 2015}, {\em 2015},~014018.

\bibitem[Torres-Herrera and Santos(2015)]{Torres2015}
Torres-Herrera, E.J.; Santos, L.F.
\newblock Dynamics at the many-body localization transition.
\newblock {\em Phys. Rev. B} {\bf 2015}, {\em 92},~014208.

\bibitem[Torres-Herrera \em{et~al.}(2016)Torres-Herrera, T\'avora, and
  Santos]{Torres2016BJP}
Torres-Herrera, E.J.; T\'avora, M.; Santos, L.F.
\newblock Survival Probability of the N\'eel State in Clean and Disordered
  Systems: An Overview.
\newblock {\em Braz. J. Phys.} {\bf 2016}, {\em 46},~239--247.

\bibitem[T\'avora \em{et~al.}()T\'avora, Torres-Herrera, and
  Santos]{TavoraARXIV}
T\'avora, M.; Torres-Herrera, E.J.; Santos, L.F.
\newblock Powerlaw Decay Exponents as Predictors of Thermalization in Many-Body
  Quantum Systems.
\newblock {\bf2016},
 arXiv:1601.05807.

\bibitem[Bohigas \em{et~al.}(1984)Bohigas, Giannoni, and Schmit]{Bohigas1984}
Bohigas, O.; Giannoni, M.J.; Schmit, C.
\newblock Characterization of Chaotic Quantum Spectra and Universality of Level
  Fluctuation Laws.
\newblock {\em Phys. Rev. Lett.} {\bf 1984}, {\em 52},~1, doi:10.1103/PhysRevLett.52.1.

\bibitem[Trotzky \em{et~al.}(2008)Trotzky, Cheinet, F\"olling, Feld,
  Schnorrberger, Rey, Polkovnikov, Demler, Lukin, and Bloch]{Trotzky2008}
Trotzky, S.; Cheinet, P.; F\"olling, S.; Feld, M.; Schnorrberger, U.; Rey,
  A.M.; Polkovnikov, A.; Demler,~E.A.; Lukin, M.D.; Bloch, I.
\newblock Time-Resolved Observation and Control of Superexchange Interactions
  with Ultracold Atoms in Optical Lattices.
\newblock {\em Science} {\bf 2008}, {\em 319},~295--299.

\bibitem[Trotzky \em{et~al.}(2012)Trotzky, Chen, Flesch, McCulloch,
  Schollw\"ock, Eisert, and Bloch]{Trotzky2012}
Trotzky, S.; Chen, Y.-A.; Flesch, A.; McCulloch, I.P.; Schollw\"ock, U.; Eisert,
  J.; Bloch, I.
\newblock Probing the relaxation towards equilibrium in an isolated strongly
  correlated one-dimensional {B}ose gas. {\em Nat. Phys.}~{\bf 2012},~{\em 8}, 325--330.

\bibitem[Kinoshita \em{et~al.}(2006)Kinoshita, Wenger, and Weiss]{kinoshita06}
Kinoshita, T.; Wenger, T.; Weiss, D.S.
\newblock A quantum {Newton's} cradle.
\newblock {\em Nature} {\bf 2006}, {\em 440},~900--903.

\bibitem[Kaufman \em{et~al.}(2016)Kaufman, M.~Eric~Tai, Rispoli, Schittko,
  Preiss, and Greiner]{Kaufman2016}
Kaufman, A.M.; Tai, M.E.; Lukin, A.; Rispoli, M.; Schittko, R.; Preiss, P.M.;
  Greiner, M. Quantum thermalization through entanglement in an isolated many-body
  system.
\newblock {\em Science} {\bf 2016}, {\em 353},~794--800.

\bibitem[Wigner(1958)]{Wigner1958}
Wigner, E.P.
\newblock On the Distribution of the Roots of Certain Symmetric Matrices.
\newblock {\em Ann. Math.} {\bf 1958}, {\em 67},~325--327.

\bibitem[Dyson(1962)]{Dyson1962}
Dyson, F.J.
\newblock The Threefold Way. Algebraic Structure of Symmetry Groups and
  Ensembles in Quantum Mechanics.
\newblock {\em J. Math. Phys.} {\bf 1962}, {\em 3},~1199--1215.

\bibitem[Mehta(1991)]{MehtaBook}
Mehta, M.L.
\newblock {\em Random Matrices}; Academic Press: Boston, MA, USA, 1991.

\bibitem[Wigner(1955)]{Wigner1955}
Wigner, E.P.
\newblock Characteristic Vectors of Bordered Matrices With Infinite Dimensions.
\newblock {\em Ann. Math.} {\bf 1955}, {\em 62},~548--564.

\bibitem[Amico \em{et~al.}(2008)Amico, Fazio, Osterloh, and Vedral]{Amico2008}
Amico, L.; Fazio, R.; Osterloh, A.; Vedral, V.
\newblock Entanglement in many-body systems.
\newblock {\em Rev. Mod. Phys.} {\bf 2008}, {\em 80},~517.

\bibitem[Lubkin(1978)]{Lubkin1978}
Lubkin, E.
\newblock Entropy of an n-system from its correlation with a k-reservoir.
\newblock {\em J. Math. Phys.} {\bf 1978}, {\em 19},~1028--1031.

\bibitem[Page(1993)]{Page1993}
Page, D.N.
\newblock Average entropy of a subsystem.
\newblock {\em Phys. Rev. Lett.} {\bf 1993}, {\em 71},~1291.

\bibitem[Erd\'elyi(1956)]{Erdelyi1956}
Erd\'elyi, A.
\newblock Asymptotic Expansions of Fourier Integrals Involving Logarithmic
  Singularities.
\newblock {\em J. Soc. Ind. Appl. Math.} {\bf 1956}, {\em 4},~38--47.

\bibitem[Urbanowski(2009)]{Urbanowski2009}
Urbanowski, K.
\newblock General properties of the evolution of unstable states at long times.
\newblock {\em Eur. Phys. J. D} {\bf 2009}, {\em 54},~25--29.

\bibitem[Khalfin(1958)]{Khalfin1958}
Khalfin, L.A.
\newblock Contribution to the decay theory of a quasi-stationary state.
\newblock {\em Sov. J. Exp. Theor. Phys.} {\bf 1958}, {\em 6},~1053--1063.

\bibitem[Muga \em{et~al.}(2009)Muga, Ruschhaupt, and del Campo]{MugaBook}
Muga, J.G.; Ruschhaupt, A.; del Campo, A.
\newblock {\em Time in Quantum Mechanics. Vol. 2}; Springer: Berlin/Heidelberg, Germany, 2009; p. 362.

\bibitem[del Campo(2011)]{Campo2011}
Del Campo, A.
\newblock Long-time behavior of many-particle quantum decay.
\newblock {\em Phys. Rev. A} {\bf 2011}, {\em 84},~012113.

\bibitem[Flambaum and Izrailev(2001)]{Flambaum2001b}
Flambaum, V.V.; Izrailev, F.M.
\newblock Entropy production and wave packet dynamics in the Fock space of
  closed chaotic many-body systems.
\newblock {\em Phys. Rev. E} {\bf 2001}, {\em 64},~036220.

\bibitem[Santos \em{et~al.}(2012{\natexlab{a}})Santos, Borgonovi, and
  Izrailev]{Santos2012PRL}
Santos, L.F.; Borgonovi, F.; Izrailev, F.M.
\newblock Chaos and Statistical Relaxation in Quantum Systems of Interacting
  Particles.
\newblock {\em Phys. Rev. Lett.} {\bf 2012}, {\em 108},~094102.

\bibitem[Santos \em{et~al.}(2012{\natexlab{b}})Santos, Borgonovi, and
  Izrailev]{Santos2012PRE}
Santos, L.F.; Borgonovi, F.; Izrailev, F.M.
\newblock Onset of chaos and relaxation in isolated systems of interacting
  spins-1/2: Energy shell approach.
\newblock {\em Phys. Rev. E} {\bf 2012}, {\em 85},~036209.

\bibitem[Berman \em{et~al.}(2004)Berman, Borgonovi, Izrailev, and
  Smerzi]{Berman2004}
Berman, G.P.; Borgonovi, F.; Izrailev, F.M.; Smerzi, A.
\newblock Irregular Dynamics in a One-Dimensional Bose System.
\newblock {\em Phys. Rev. Lett.} {\bf 2004}, {\em 92},~030404.

\bibitem[Haldar \em{et~al.}(2016)Haldar, Chavda, Vyas, and Kota]{Haldar2016}
Haldar, S.K.; Chavda, N.D.; Vyas, M.; Kota, V.K.B.
\newblock Fidelity decay and entropy production in many-particle systems after
  random interaction quench.
\newblock {\em J. Stat. Mech. Theory Exp.} {\bf
  2016}, {\em 2016},~043101.

\bibitem[Srednicki(1996)]{Srednicki1996}
Srednicki, M.
\newblock Thermal fluctuations in quantized chaotic systems.
\newblock {\em J. Phys. A Math. Gen.} {\bf 1996}, {\em 29},~L75.

\bibitem[Srednicki(1999)]{Srednicki1999}
Srednicki, M.
\newblock The approach to thermal equilibrium in quantized chaotic systems.
\newblock {\em J. Phys. A Math. Gen.} {\bf 1999}, {\em 32},~1163.

\bibitem[Reimann(2008)]{Reimann2008}
Reimann, P.
\newblock Foundation of Statistical Mechanics under Experimentally Realistic
  Conditions.
\newblock {\em Phys. Rev. Lett.} {\bf 2008}, {\em 101},~190403.

\bibitem[Short(2011)]{Short2011}
Short, A.J.
\newblock Equilibration of quantum systems and subsystems.
\newblock {\em New J. Phys.} {\bf 2011}, {\em 13},~053009.

\bibitem[Gramsch and Rigol(2012)]{Gramsch2012}
Gramsch, C.; Rigol, M.
\newblock Quenches in a quasidisordered integrable lattice system: Dynamics and
  statistical description of observables after relaxations.
\newblock {\em Phys. Rev. A} {\bf 2012}, {\em 86},~053615.

\bibitem[Venuti and Zanardi(2013)]{Venuti2013}
Venuti, L.C.; Zanardi, P.
\newblock Gaussian equilibration.
\newblock {\em Phys. Rev. E} {\bf 2013}, {\em 87},~012106.

\bibitem[He \em{et~al.}(2013)He, Santos, Wright, and Rigol]{HeSantos2013}
He, K.; Santos, L.F.; Wright, T.M.; Rigol, M.
\newblock Single-particle and many-body analyses of a quasiperiodic integrable
  system after a quench.
\newblock {\em Phys. Rev. A} {\bf 2013}, {\em 87},~063637.

\bibitem[Zangara \em{et~al.}(2013)Zangara, Dente, Torres-Herrera, Pastawski,
  Iucci, and Santos]{Zangara2013}
Zangara, P.R.; Dente, A.D.; Torres-Herrera, E.J.; Pastawski, H.M.; Iucci, A.;
  Santos, L.F.
\newblock Time fluctuations in isolated quantum systems of interacting
  particles.
\newblock {\em Phys. Rev. E} {\bf 2013}, {\em 88},~032913.

\bibitem[Kiendl and Marquardt()]{KiendlARXIV}
Kiendl, T.; Marquardt, F.
\newblock Many-particle dephasing after a quench. {\bf 2016}, arXiv:1603.01071.

\bibitem[Chirikov(1986)]{Chirikov1986}
Chirikov, B.V.
\newblock Transient chaos in quantum and classical mechanics.
\newblock {\em Found. Phys.} {\bf 1986}, {\em 16},~39--49.

\bibitem[Chirikov(1997)]{Chirikov1997}
Chirikov, B.V.
\newblock Linear and nonlinear dynamical chaos.
\newblock {\em Open Sys. Inf. Dyn.} {\bf 1997}, {\em 4},~241--280.

\bibitem[Robinett(2004)]{Robinett2004}
Robinett, R.W.
\newblock Quantum wave packet revivals.
\newblock {\em Phys. Rep.} {\bf 2004}, {\em 392},~1--119.

\bibitem[Polkovnikov(2011)]{Polkovnikov2011}
Polkovnikov, A.
\newblock Microscopic diagonal entropy and its connection to basic
  thermodynamic relations.
\newblock {\em Ann.~Phys.} {\bf 2011}, {\em 326},~486--499.

\bibitem[Santos \em{et~al.}(2011)Santos, Polkovnikov, and Rigol]{Santos2011PRL}
Santos, L.F.; Polkovnikov, A.; Rigol, M.
\newblock Entropy of isolated quantum systems after a quench.
\newblock {\em Phys. Rev.~Lett.} {\bf 2011}, {\em 107},~040601.

\bibitem[Santos \em{et~al.}(2012)Santos, Polkovnikov, and
  Rigol]{Santos2012PRER}
Santos, L.F.; Polkovnikov, A.; Rigol, M.
\newblock Weak and strong typicality in quantum systems.
\newblock {\em Phys. Rev. E} {\bf 2012}, {\em 86},~010102.

\bibitem[Deutsch(1991)]{Deutsch1991}
Deutsch, J.M.
\newblock Quantum statistical mechanics in a closed system.
\newblock {\em Phys. Rev. A} {\bf 1991}, {\em 43},~2046.

\bibitem[Srednicki(1994)]{Srednicki1994}
Srednicki, M.
\newblock Chaos and quantum thermalization.
\newblock {\em Phys. Rev. E} {\bf 1994}, {\em 50},~888.

\bibitem[Rigol \em{et~al.}(2008)Rigol, Dunjko, and Olshanii]{Rigol2008}
Rigol, M.; Dunjko, V.; Olshanii, M.
\newblock Thermalization and its mechanism for generic isolated quantum
  systems.
\newblock {\em Nature} {\bf 2008}, {\em 452},~854--858.

\bibitem[Rigol(2009{\natexlab{a}})]{rigol09STATa}
Rigol, M.
\newblock Breakdown of thermalization in finite one-dimensional systems.
\newblock {\em Phys. Rev. Lett.} {\bf 2009}, {\em 103},~100403.

\bibitem[Rigol(2009{\natexlab{b}})]{rigol09STATb}
Rigol, M.
\newblock Quantum quenches and thermalization in one-dimensional fermionic
  systems.
\newblock {\em Phys. Rev. A} {\bf 2009}, {\em 80},~053607.

\bibitem[Torres-Herrera and Santos(2013)]{Torres2013}
Torres-Herrera, E.J.; Santos, L.F.
\newblock Effects of the interplay between initial state and {H}amiltonian on
  the thermalization of isolated quantum many-body systems.
\newblock {\em Phys. Rev. E} {\bf 2013}, {\em 88},~042121.

\bibitem[He and Rigol(2013)]{He2013}
He, K.; Rigol, M.
\newblock Initial-state dependence of the quench dynamics in integrable quantum
  systems. III. Chaotic states.
\newblock {\em Phys. Rev. A} {\bf 2013}, {\em 87},~043615.

\bibitem[Santos and Rigol(2010)]{Santos2010PRE}
Santos, L.F.; Rigol, M.
\newblock Onset of quantum chaos in one-dimensional bosonic and fermionic
  systems and its relation to thermalization.
\newblock {\em Phys. Rev. E} {\bf 2010}, {\em 81},~036206.

\bibitem[Rigol and Santos(2010)]{RigolSantos2010}
Rigol, M.; Santos, L.F.
\newblock Quantum chaos and thermalization in gapped systems.
\newblock {\em Phys. Rev. A} {\bf 2010}, {\em 82},~011604.

\bibitem[Santos and Rigol(2010)]{Santos2010PREb}
Santos, L.F.; Rigol, M.
\newblock Localization and the effects of symmetries in the thermalization
  properties of one-dimensional quantum systems.
\newblock {\em Phys. Rev. E} {\bf 2010}, {\em 82},~031130.

\bibitem[Neuenhahn and Marquardt(2012)]{neuenhahn_marquardt_12}
Neuenhahn, C.; Marquardt, F.
\newblock Thermalization of interacting fermions and delocalization in Fock
  space.
\newblock {\em Phys. Rev. E} {\bf 2012}, {\em 85},~060101.

\bibitem[Jensen and Shankar(1985)]{Jensen1985}
Jensen, R.V.; Shankar, R.
\newblock Statistical Behavior in Deterministic Quantum Systems with Few
  Degrees of Freedom.
\newblock {\em Phys. Rev. Lett.} {\bf 1985}, {\em 54},~1879.

\bibitem[Dyson(1962)]{Dyson1962b}
Dyson, F.J.
\newblock Statistical Theory of the Energy Levels of Complex Systems. I.
\newblock {\em J. Math. Phys.} {\bf 1962}, {\em 3},~140--156.

\bibitem[French and Wong(1970)]{French1970}
French, J.B.; Wong, S.S.M.
\newblock Validity of random matrix theories for many-particle systems.
\newblock {\em Phys. Lett. B} {\bf 1970}, {\em 33},~449--452.

\bibitem[Bohigas and Flores(1971)]{Bohigas1971}
Bohigas, O.; Flores, J.
\newblock Two-body random Hamiltonian and level density.
\newblock {\em Phys. Lett. B} {\bf 1971}, {\em 34},~261--263.

\bibitem[Brody \em{et~al.}(1981)Brody, Flores, French, Mello, Pandey, and
  Wong]{Brody1981}
Brody, T.A.; Flores, J.; French, J.B.; Mello, P.A.; Pandey, A.; Wong, S.S.M.
\newblock Random-matrix physics: Spectrum and strength fluctuations.
\newblock {\em Rev. Mod. Phys} {\bf 1981}, {\em 53},~385.

\bibitem[Mirlin \em{et~al.}(1996)Mirlin, Fyodorov, Dittes, Quezada, and
  Seligman]{Mirlin1996}
Mirlin, A.D.; Fyodorov, Y.V.; Dittes, F.M.; Quezada, J.; Seligman, T.H.
\newblock Transition from localized to extended eigenstates in the ensemble of
  power-law random banded matrices.
\newblock {\em Phys. Rev. E} {\bf 1996}, {\em 54},~3221.

\bibitem[Santos(2004)]{Santos2004}
Santos, L.F. Integrability of a disordered {H}eisenberg spin-1/2 chain. 
{\em J. Phys. A Math. Gen.}~{\bf 2004},~{\em 37},~4723.

\bibitem[Rufeil-Fiori and Pastawski(2006)]{Fiori2006}
Rufeil-Fiori, E.; Pastawski, H.M.
\newblock Non-Markovian decay beyond the Fermi Golden Rule: Survival collapse
  of the polarization in spin chains.
\newblock {\em Chem. Phys. Lett.} {\bf 2006}, {\em 420},~35--41.

\bibitem[Rufeil-Fiori and Pastawski(2009)]{Fiori2009}
Rufeil-Fiori, E.; Pastawski, H.M.
\newblock Survival probability of a local excitation in a non-Markovian
  environment: Survival collapse, Zeno and anti-Zeno effects.
\newblock {\em Phys. B Condens. Matter} {\bf 2009}, {\em 404},~2812--2815.

\bibitem[Lawrence(2002)]{Lawrence2002}
Lawrence, J.
\newblock Nonexponential decay at late times and a different Zeno paradox.
\newblock {\em J. Opt. B} {\bf 2002}, {\em 4},~S446.

\bibitem[Sidje(1998)]{Sidje1998}
Sidje, R.B.
\newblock Expokit: A software package for computing matrix exponentials.
\newblock {\em ACM Trans. Math. Softw.} {\bf 1998}, {\em 24},~130--156.

\bibitem[Exp()]{Expokit}
{Expokit}. Available online: http://www.maths.uq.edu.au/expokit/ (accessed on 29 September 2016).

\end{thebibliography}
\end{document}